\documentclass[12pt, oneside]{article}   	
\usepackage[T1]{fontenc}
\usepackage{lmodern}
\usepackage{geometry}                		
\usepackage{graphicx}				
\usepackage{amsmath}		
\usepackage{amssymb,amsfonts,amsthm}
\usepackage{mathtools,cool}
\usepackage{braket}

\title{Four Dimensional Anomaly-Free Twistor String}
\author{Christian Kunz \\ \small{\textit{E-mail:} \href{mailto:kunz.christian.321@gmail.com}{kunz.christian.321@gmail.com}}}
\usepackage{plain}
\usepackage{hyperref}

\newcommand{\ud}{\mathrm{d}}
\numberwithin{equation}{section}

\begin{document}
  \maketitle
  \tableofcontents
  
\begin{abstract}
  This paper presents a covariant twistor string model in four dimensions coupled to two-dimensional worldsheet gravity with an anomaly-free BRST charge. The model contains two twistors and two fermionic bi-spinors with incidence related gauge symmetries and an SU(2) gauge symmetry between the twistors. It is shown that the model leads to familiar gravitational N$^k$MHV amplitudes and has modular-invariant one-loop amplitudes.
\end{abstract}

\section{Introduction}
     Twistor space has been successfully used to describe (anti-)self-dual gravity with help of the Penrose transform
\cite {Penrose:1986} . With somewhat less practical usefulness, a correspondence between full complex conformal gravity and certain complex contact structures in ambitwistor space has been established
\cite {LeBrun:1991}. Penrose originally envisioned twistors to be the origin for a quantization of gravity, but this programme had limited success. Nowadays there are more popular avenues for quantum gravity, like superstrings, loop quantum gravity, asymptotically safe spacetimes, and more. Twistor string\footnote{The twistor string revolution started with Witten's seminal paper \cite{Witten:2004}.} actions have been considered for computing (super-)Yang-Mills and supergravity scattering amplitudes, but not really as basis for a full-fledged quantum gravity that includes
matter.\\

The model presented here is a string model with target space being 4-dimensional twistor space with its dual, it is coupled to 2-dimensional worldsheet gravity, and it has fermionic gauge symmetries, which do not seem to arise from conventional supersymmetry in superspace, at least not in an obvious way. The spectrum includes a number of  lowest level modes, in both the Neveu-Schwarz and the Ramond sector. Most importantly, the model has an anomaly-free BRST charge\footnote{The N=8 twistor string theory of Skinner%
\cite {Skinner:2013} has an anomaly-free BRST charge as well, but is not coupled to worldsheet gravity.}.
N$^k$MHV tree scattering amplitudes have the familiar generalized Hodges form
\cite {Geyer:2014}, and one-loop scattering amplitudes can be shown to exhibit modular invariance without need for a GSO-type projection. Modular invariance often is a good starting point to prove consistency and unitarity, but for now this remains a speculation.\\

In section 2 the model is presented, referring to Appendix A for calculating the commutators of the currents, proving that the resulting BRST charge Q is anomaly free and
nilpotent fulfilling $Q ^ {2} =0$, and showing that the theory has ground states in both the Neveu-Schwarz and Ramond sector.

In section 3 vertex operators are presented and used to calculate scattering amplitudes up to one-loop level. It is shown that the tree N$^k$MHV amplitudes have the familiar form, and that the one-loop amplitudes are modular invariant.

The last section contains final remarks and outlook.

\section{The Twistor String Model}
\label{Model}
  The Euclidean worldsheet actions of the twistor string model presented here is:
  \begin{multline}
  S_0 = \frac{1}{2 \pi} \int\mathop{}\negthickspace{\ud^2\mathop{}\negthickspace z} \left\{
  \frac{1}{2} \sum_{i=1}^2 \left( Y_i\overline\partial{Z_i} - Z_i \overline\partial{Y_i} + \Theta_i\overline\partial{\Psi_i}  + \Psi_i\overline\partial{\Theta_i} \right)
  \right.\\
  + \sum_{i,j=1}^2 \lambda_i a_{1 i j} \tilde{\phi}_j
  + \sum_{i,j=1}^2 \tilde{\lambda}_i a_{2 i j} \phi_j 
  + \sum_{i,j=1}^2\tilde{\lambda}_i b_{i j} \lambda_j
  + \left(Y_1 Y_2 \right) \mathop{}\negthickspace \vec{c} \, \cdotp \vec{\tau}  \mathop{}\negthickspace \begin{pmatrix} Z_1 \\ Z_2 \end{pmatrix} 
  \left. \vphantom{\sum_i} \right\}
  \label{action}
  \end{multline}
  \begin{multline*}
  \hat{S}_0 = \frac{1}{2 \pi} \int\mathop{}\negthickspace{\ud^2\mathop{}\negthickspace z}  \left\{
  \frac{1}{2} \sum_{i=1}^2 \left( \hat{Y}_i \partial{\hat{Z}_i} - \hat{Z}_i \partial{\hat{Y}_i} + \hat{\Theta}_i \partial{\hat{\Psi}_i}  + \hat{\Psi}_i \partial{\hat{\Theta}_i} \right)
  \right.\\
  + \sum_{i,j=1}^2 \hat{\lambda}_i \hat{a}_{1 i j} \tilde{\hat{\phi}}_j
  + \sum_{i,j=1}^2 \tilde{\hat{\lambda}}_i \hat{a}_{2 i j} \hat{\phi}_j 
  + \sum_{i,j=1}^2\tilde{\hat{\lambda}}_i \hat{b}_{i j} \hat{\lambda}_j
  + \left(\hat{Y}_1 \hat{Y}_2 \right) \mathop{}\negthickspace \vec{\hat{c}} \, \cdotp \vec{\tau}  \mathop{}\negthickspace \begin{pmatrix} \hat{Z}_1 \\ \hat{Z}_2 \end{pmatrix} 
  \left. \vphantom{\sum_i} \right\} 
  \end{multline*}
 where $Z_i/\hat{Z}_i$ are left/right moving twistors, $Y_i/\hat{Y}_i$ dual twistors, $\Psi_i/\hat{\Psi}_i$ fermionic bi-spinors, and $\Theta_i/\hat{\Theta}_i$ fermionic dual bi-spinors, each of the left moving entities with components:
 \begin{equation*}
 Z_i = \binom{\lambda_{i \alpha_i}}{\mu_i^{\dot{\alpha}_i}}, Y_i = \binom{\tilde{\mu}_i^{\alpha_i}}{\tilde{\lambda}_{i \dot{\alpha}_i}},
 \Psi_i = \binom{\phi_{i \alpha_i}}{\psi_i^{\dot{\alpha}_i}}, \Theta_i = \binom{\tilde{\psi}_i^{\alpha_i}}{\tilde{\phi}_{i \dot{\alpha}_i}},
 \end{equation*}
 similarly for the right moving entities,
 and $\tau^i$ are the Pauli matrices.
 
 Further,
 $a_{1 i j}\! =\! a_{1 i j}^{\alpha_i \dot{\alpha}_j}$ (fermionic), 
 $a_{2 i j}\! =\! a_{2 i j}^{\dot{\alpha}_i \alpha_j}$ (fermionic),
 $b_{i j}\! =\! b_{i j}^{\dot{\alpha}_i \alpha_j}$ (bosonic), and $\vec{c}$ (bosonic) are Lagrange multipliers, and the same for the $\hat{a}_{1 i j}$, $\hat{a}_{1 i j}$,
 $\hat{b}_{i j}$, and $\vec{\hat{c}}$.\\[1pt]

 Similar to the NSR superstring, the action \eqref{action} is the conformal gauge result from
  coupling to worldsheet gravity with reparameterization invariance and has the following additional gauge symmetries:
  
  \begin{enumerate}
  
  \item $4 \mathop{}\negthickspace \times \mathop{}\negthickspace 4$-fold bosonic: for $i,j \in \{1,2\}$
  \begin{equation*}
  \begin{split}
  & \hphantom{HHHH} \delta \mu_i^{\dot{\alpha}} = \varepsilon^{\dot{\alpha} \alpha} \lambda_{j \alpha}, \\  
  & \hphantom{HHHH} \delta \tilde\mu_j^{\alpha} = - \varepsilon^{\dot{\alpha} \alpha} \tilde\lambda_{i {\dot\alpha}},\\  
  & \hphantom{HHHH} \delta b_{i j}^{\dot{\alpha} \alpha} = -\overline\partial \varepsilon^{\dot{\alpha} \alpha} + \frac{1}{2} \left((-1)^i \mathop{}\negthickspace - \mathop{}\negthickspace (-1)^j)\right) \mathop{}\negthickspace c_3\, \varepsilon^{\dot{\alpha} \alpha} ,\\
  & \hphantom{HHHH} \delta b_{i\: 3-j}^{\dot{\alpha} \alpha} =  \frac{1}{2} \left( c_1 -i (-1)^j \mathop{}\negthickspace c_2 \right) \varepsilon^{\dot{\alpha} \alpha} \\
  & \hphantom{HHHH} \delta b_{3-i\: j}^{\dot{\alpha} \alpha} =  - \frac{1}{2} \left( c_1 -i (-1)^i \mathop{}\negthickspace c_2 \right) \varepsilon^{\dot{\alpha} \alpha}.
  \end{split}
  \end{equation*}
  
  \item $4 \mathop{}\negthickspace \times \mathop{}\negthickspace 4$-fold fermionic: for $i,j \in \{1,2\}$
  \begin{equation*}\ 
  \begin{split}
  & \delta \tilde\mu_i^{\alpha} = \varepsilon^{\alpha \dot{\beta}} \tilde\phi_{j {\dot\beta}},\\ 
  & \delta \psi_j^{\dot{\beta}} = \varepsilon^{\alpha \dot{\beta}} \lambda_{i \alpha},\\ 
  & \delta a_{1 i j}^{\alpha \dot{\beta}} = \overline\partial \varepsilon^{\alpha \dot{\beta}} + \frac{1}{2} (-1)^i  \mathop{}\negthickspace c_3\, \varepsilon^{\alpha \dot{\beta}} ,\\
  & \delta a_{1\, 3-i\, j}^{\alpha \dot{\beta}} = - \frac{1}{2} \left( c_1 + i (-1)^i \mathop{}\negthickspace c_2 \right) \varepsilon^{\alpha \dot{\beta}} .
  \end{split}
  \end{equation*}
  
  \item $4 \mathop{}\negthickspace \times \mathop{}\negthickspace 4$-fold fermionic: for $i,j \in \{1,2\}$
  \begin{equation*}\ 
  \begin{split}
  & \delta \mu_i^{\dot{\alpha}} = \varepsilon^{\dot{\alpha} \beta} \phi_{j \beta},\\ 
  & \delta \tilde\psi_j^{\beta} = - \varepsilon^{\dot{\alpha} \beta} \tilde\lambda_{i \dot{\alpha}},\\ 
  & \delta a_{2 i j}^{\dot{\alpha} \beta} = - \overline\partial \varepsilon^{\dot{\alpha} \beta} + \frac{1}{2} (-1)^i  \mathop{}\negthickspace c_3\, \varepsilon^{\dot{\alpha} \beta} ,\\
  & \delta a_{2\, 3-i\, j}^{\dot{\alpha} \beta} =  - \frac{1}{2} \left( c_1 - i (-1)^i \mathop{}\negthickspace c_2 \right) \varepsilon^{\dot{\alpha} \beta} .
  \end{split}
  \end{equation*}
  
  \item $SU(2)$ between the two twistors:
  \begin{equation*}
  \begin{split}
  & \hphantom{H}\ \delta \binom{Z_1}{Z_2} = \vec{\varepsilon}  \cdotp \vec{\tau} \binom{Z_1}{Z_2},\\
  & \hphantom{H}\ \delta \left(Y_1 Y_2 \right) = -\left(Y_1 Y_2 \right) \vec{\varepsilon}  \cdotp \vec{\tau},\\
  & \hphantom{H}\ \delta\, \vec{c} = - \overline\partial\, \vec{\varepsilon} + i ( \vec{\varepsilon} \times \vec{c} ),\\
  & \hphantom{H}\ \delta \left(a_{1 1 j} a_{1 2 j} \right) = -\left(a_{1 1 j} a_{1 2 j} \right) \vec{\varepsilon}  \cdotp \vec{\tau},\\
  & \hphantom{H}\ \delta \binom{a_{2 1 j}}{a_{2 2 j}} = \vec{\varepsilon}  \cdotp \vec{\tau} \binom{a_{2 1 j}}{a_{2 2 j}},\\
  & \hphantom{H}\ \delta b_{i j} = \vec{\varepsilon}  \cdotp\! \left( \vec{\tau} \binom{b_{1 j}}{b_{2 j}} \right)_{\!i}\! - \left( \left(b_{i 1} b_{i 2} \right) \vec{\tau} \right)_j\!  \cdotp\! \vec{\varepsilon} .
  \end{split}
  \end{equation*}
  
  \end{enumerate}
  
  Of course, there are the equivalent symmetries for the $\langle \hat{Z}_i, \hat{Y}_i, \hat{\Psi}_i, \hat{\Theta}_i, \hat{a}_{1 i  j}, \hat{a}_{2 i j},\! \hat{b}_{i j}, \vec{\hat{c}} \rangle$ (right-moving) entities, with $\overline\partial$ replaced by $\partial$.
  
  Note also that the action does not have a Lagrange multiplier for the scaling symmetry $Z_i \rightarrow \lambda Z_i$,  $Y_i \rightarrow (1 / \lambda) Y_i$, $\Psi_i \rightarrow \lambda \Psi_i$, $\Theta_i \rightarrow (1 / \lambda) \Theta_i$, i.e. this global symmetry is not gauged and the target space includes but is not restricted to ambitwistors.
  
  Further, because some of the gauge symmetries involve just the spinor parts of the twistors and not the full twistors, the model breaks full conformal invariance by fixing the infinity twistor \cite{Penrose:1986}.\\[1pt]
  
  BRST quantization is presented in the Appendix~\ref{BRST}, where it is shown that the model has an anomaly-free BRST charge $Q$ and non-trivial physical ground states in both the Neveu-Schwarz (NS) and Ramond (R) sector.\\
  
  At this stage it is not clear yet which types of string theories are of physical interest: oriented open (and closed) chiral strings with modes moving in a single direction, unoriented open (and closed) strings,  and/or oriented closed strings with modes moving in both directions (adding the two actions $S_0$ and $\hat{S}_0$ together).

\section{Vertex Operators and Scattering Amplitudes}
  \label{VertexOps}
  
  For the sake of simplicity we limit ourselves in this section to left-moving open-closed strings.
  In analogy to the scattering amplitude for N vertex operators in the NSR superstring theory (see e.g. \cite{Polyakov:2005}) the amplitude here at genus $g$ in picture number
  $\mp1$ for the bosonic/fermionic NS sector and $\mp1/2$ for the bosonic/fermionic R sector is given by  
  \begin{multline*}
  < \!\! V_1(z_1), \dots, V_N(z_N) \!\! > = \\ \int \! 
  \prod_{i=1}^{M} \!  \mathrm{d}m_i \!  \prod_{s=1}^{S}  \mathrm{d}\vec{n}_s \! \!
  \prod_{k,l=1 \atop {\alpha, \dot{\alpha}=1}}^2 \! \! \prod_{s=1}^{S_{kl}} \! \mathrm{d}m_{skl}^{\alpha \dot{\alpha}} \!\!
  \prod_{j,k,l=1 \atop {\alpha, \dot{\alpha}=1}}^2 \! \prod_{a=1}^{P_{jkl}} \mathrm{d}\theta_{ajkl}^{\alpha \dot{\alpha}} 
  \int \prod_{r=1}^2 (\mathrm{D}Z_r  \mathrm{D}Y_r \mathrm{D} \Psi_r   \mathrm{D} \Theta_r \mathrm{D} [\text{ghosts}] \\
  \exp [- S_1 + m_i \!  \braket{ \xi^i | T } \!  + \vec{n}_s \braket{  \vec{\eta^s} | \vec{H} }
  + m_{skl}^{\alpha \dot{\alpha}} \!  \braket{ \xi^{skl}_{\alpha \dot{\alpha}} | F_{kl \, \alpha \dot{\alpha}} } \! 
  + \theta_{ajkl}^{\alpha \dot{\alpha}} \!  \braket{ \chi^{ajkl}_{\alpha \dot{\alpha}} | G_{jkl \, \alpha \dot{\alpha}} }] \\
  \prod_{i=1}^{M} \!  \delta ( \braket{\xi^i | b} ) \! \prod_{s=1}^{S} \!  \delta ( \braket{\vec{\eta^s} | \vec{h}} ) \!\!\!
  \prod_{k,l=1 \atop {\alpha, \dot{\alpha}=1}}^2 \! \prod_{s=1}^{S_{kl}} \!  \delta ( \braket{\xi^{skl}_{\alpha \dot{\alpha}} | f_{kl \, \alpha \dot{\alpha}} })
  \!\! \! \prod_{j,k,l=1 \atop {\alpha, \dot{\alpha}=1}}^2 \! \prod^{P_{jkl}}_{a=1} \! \delta (\braket{ \chi^{ajkl}_{\alpha \dot{\alpha}} | \beta_{jkl \, \alpha \dot{\alpha}} })  V_1(\!z_1\!), \dots, V_N(\!z_N\!) \, ,
  \end{multline*}
  where in the exponential summation over repeated indices is assumed, $S_1$ is the gauge-fixed action \eqref{gf_action},   
  \begin{align*}
  T & = \{Q, b\},                          & \vec{H} & = \{Q, \vec{h}\},  
  & F_{kl \, \alpha \dot{\alpha}} & = \{Q, f_{kl \, \alpha \dot{\alpha}}\},
  & G_{jkl \, \alpha \dot{\alpha}} & = [Q, \beta_{jkl \, \alpha \dot{\alpha}}]
  \end{align*}
  
  are the full currents including ghosts, $Q$ is the BRST charge, $(m_i, \vec{n}_s, m_{skl}^{\alpha \dot{\alpha}}; \theta_{ajkl}^{\alpha \dot{\alpha}})$ are the holomorphic even and odd coordinates in the moduli space,
  $(\xi^i,  \vec{\eta^s}, \xi^{skl}_{\alpha \dot{\alpha}}; \chi^{ajkl}_{\alpha \dot{\alpha}})$ are their dual  Beltrami differentials, the $\braket{ \cdots | \cdots }$ symbol stands for the scalar product in the Hilbert space, and the delta-functions 
  $\delta (\braket{\chi^{ajkl}_{\alpha \dot{\alpha}} | \beta_{jkl \, \alpha \dot{\alpha}} } )$, $\delta(\braket{\xi^i | b } ) = \braket{ \xi^i | b }$, $\delta( \braket{\vec{\eta^s} | \vec{h} } )$, and $\delta ( \braket{ \xi^{skl}_{\alpha \dot{\alpha}} | f_{kl \, \alpha \dot{\alpha}} } )$ are needed to insure that the basis in the moduli space is normal to variations along the conformal gauge slices. The dimensions of the moduli spaces are related to the number $N$ of the vertex operators present and the various numbers of NS and R punctures of the bosonic and fermionic excitations they represent. Dimensions $M$ and $S$ are clear for the $b-c$ and the $\vec{h}-\vec{g}$ ghosts, resp.: 
 \begin{align*} 
 M & = 3(g-1) + N\\
 S & = g - 1 + N
 \end{align*}
 but $S_{kl}$ and $P_{jkl}$ which can be smaller than $S$ depend on the details of the scattering situation.
   
  We introduce a couple of fixed vertex operators in the NS sector that commute with the BRST charge $Q$ \eqref{q_charge}:
  
  \begin{equation*}
  \begin{split}
  \mathcal{V}_{iq} & = c g_1 g_2 g_3 \!\!\! \prod_{k,l=1 \atop {\alpha, \dot{\alpha}=1}}^2 \!\!\! \left(\!\!e_{kl}^{\alpha \dot{\alpha}} \! \prod_{j=1}^2 \! \delta(\gamma_{jkl}^{\alpha \dot{\alpha}})\!\!\right) \!
  \int \!\! \frac{\mathrm{d}t} {t^3}\! : \!\delta^2\!(\rho_i \!-\! t \lambda_q(z))  \mathrm{e}^{it \tilde{\rho}_i \mu_q(z) } \! : \\
  \tilde{\mathcal{V}}_{iq} & = c g_1 g_2 g_3 \!\!\! \prod_{k,l=1 \atop {\alpha, \dot{\alpha}=1}}^2 \!\!\! \left(\!\!e_{kl}^{\alpha \dot{\alpha}}  \! \prod_{j=1}^2 \! \delta(\gamma_{jkl}^{\alpha \dot{\alpha}})\!\!\right) \!
  \int \!\! \frac{\mathrm{d}t} {t^3} \! : \! \delta^2\!(\tilde{\rho}_i \!-\! t \tilde{\lambda}_q(z)) \mathrm{e}^{it \tilde{\mu}_q(z) \rho_i}\! :
  \end{split}
  \end{equation*}
  
  Note that $\rho_i$ and $\tilde{\rho}_i$ do not have a second subscript $q$ because of the $SU(2)$ symmetry of the bosonic spinors. Also we did not include,
  as in \cite{Geyer:2014}, a factor like\\
  \hspace*{50pt}$\delta^2\!(\eta_{ir} \!-\! t \phi_r(z)) \mathrm{e}^{it \tilde{\eta}_{ir} \psi_r(z)}$ in $\tilde{V}_{iq}$ or\\
  \hspace*{50pt}$\delta^2\!(\tilde{\eta}_{ir} \!-\! t \tilde{\phi}_r(z)) \mathrm{e}^{it \tilde{\psi}_r(z) \eta_{ir} }$ in $\tilde{\mathcal{V}}_{iq}$\\
  because we  are not working in a superspace so that supermomenta $\eta, \tilde{\eta}$ do not make sense here.\\
  
  What about the Ramond sector? Because each bi-spinor or twistor consists of 4 complex fields, fermionic spinor fields and bosonic twist fields that create or annihilate a cut for 
  the bi-spinor and twistor fields, resp.,  have conformal weight $\frac{1}{2}$, i.e. behave like ordinary spinor fields.
  Therefore, Ramond vertex operators of conformal weight 1 can easily be generated by just multiplying one of them with a bi-spinor or twistor field, i.e. there is no need for ghost spinor operators. Actually many Ramond ground states are even in the number of oscillators creating them, i.e. there is no need to create or annihilate cuts for such states and the above vertex operators can be used as well. Ultimately, which sector to use is determined by the spin structure of the worldsheet.\\
  
  In order to show that the model leads to familiar N$^k$MHV amplitudes, we make some simplifications. First we limit external excitations to the bosonic sector and use the fact that for tree amplitudes the $SU(2)$ symmetry between the twistors can be taken care of by just considering one of the two twistors, omitting the index. Then we can disregard the $\vec{g}$, $\vec{h}$, $\gamma$, and $\beta$ ghosts and the vertex operators simplify to
  \begin{equation*}
  \begin{split}
  \mathcal{V}_i & = c \!\!\! \prod_{\alpha, \dot{\alpha}=1}^2 \!\! e_{\cdot\cdot}^{\alpha \dot{\alpha}} 
  \int \!\! \frac{\mathrm{d}t} {t^3} : \delta^2\!(\rho_i \!-\! t \lambda(z)) \mathrm{e}^{it \tilde{\rho}_i \mu(z)} : \\
  \tilde{\mathcal{V}}_i & = c \!\!\! \prod_{\alpha, \dot{\alpha}=1}^2 \!\! e_{\cdot\cdot}^{\alpha \dot{\alpha}}
  \int \!\! \frac{\mathrm{d}t} {t^3} : \delta^2\!(\tilde{\rho}_i \!-\! t \tilde{\lambda}(z)) \mathrm{e}^{it \tilde{\mu}(z) \rho_i} :
  \end{split}
  \end{equation*}
  
  Further we notice that for the vertex operator $\mathcal{V}_i$ most of the ghosts decouple except
  $\tilde{E}_i^{\alpha}(z) = \tilde{\rho}_{i \dot{\alpha}} e_{\cdot\cdot}^{\alpha \dot{\alpha}}(z)$ and
  $\tilde{F}_{i \, \alpha}(z) = \tilde{\rho}^{i \dot{\alpha}} f_{\cdot\cdot \, \alpha \dot{\alpha}}(z)$ (due to the term in the exponential), projected into the direction of $\rho_i$ 
  (due to the delta function) with help of a projection operator $\Pi_{(\rho_i\!) \alpha}^{\beta}$,  and  for $\tilde{\mathcal{V}}_i$ most of the ghosts except
  $E_i^{\dot{\alpha}}(z) = \rho_{i \alpha} e_{\cdot\cdot}^{\alpha \dot{\alpha}}(z)$ and $F_{i \, \dot{\alpha}}(z) = \rho^{i \alpha} f_{\cdot\cdot \, \alpha \dot{\alpha}}(z)$, 
  projected into the direction of $\tilde{\rho}_i$ with help of a projection operator $\tilde{\Pi}_{(\tilde{\rho}_i\!) \dot{\alpha}}^{\dot{\beta}}$, 
  such that the associated integrated vertex operators take the form:  
  \begin{equation}
  \begin{split}
  & V_i = \int \!\! \frac{\mathrm{d}t} {t^2} \mathrm{d}z : ([\tilde{\lambda}(z) \tilde{\rho}_i] - i\,t \tilde{E}_i^{\alpha}(z) \Pi_{(\rho_i\!) \alpha}^{\beta} \tilde{F}_{i \, \beta}(z)) \, \delta^2\!(\rho_i \!-\! t \lambda(z)) \,  \mathrm{e}^ {it [\tilde{\rho}_i \mu(z)] } : \, ,\\
  & \tilde{V}_i = \int \!\! \frac{\mathrm{d}t} {t^2} \mathrm{d}z : (\braket{\lambda(z) \rho_i } + i\,t E_i^{\dot{\alpha}}(z)  \tilde{\Pi}_{(\tilde{\rho}_i\!) \dot{\alpha}}^{\dot{\beta}} F_{i \, \dot{\beta}}(z) ) \, \delta^2\!(\tilde{\rho}_i \!-\! t \tilde{\lambda}(z)) \, \mathrm{e}^{it \braket{\tilde{\mu}(z) \rho_i } } : \\
  \text{with}\\
  & \{ E_i^{\dot{\gamma}}(z) \tilde{\Pi}_{\dot{\gamma}}^{\dot{\alpha}}, \tilde{\Pi}_{\dot{\beta}}^{\dot{\kappa}} F_{j \, \dot{\kappa}}(w) \} = 
  \tilde{\Pi}_{\dot{\beta}}^{\dot{\alpha}} \, \delta(z - w) \braket{\rho_i \rho_j } \\
  & \{ \tilde{E}_i^{\gamma}(z) \Pi_{\gamma}^{\alpha}, \Pi_{\beta}^{\kappa} \tilde{F}_{j \, \kappa}(w) \} = \Pi_{\beta}^{\alpha} \, \delta(z - w) \, [\tilde{\rho}_i \tilde{\rho}_j ]
  \end{split}
  \label{GRvertex}
  \end{equation}
  Here we used $\tilde{\Pi}_{\dot{\gamma}}^{\dot{\alpha}} \tilde{\Pi}_{\dot{\beta}}^{\dot{\gamma}}  = \tilde{\Pi}_{\dot{\beta}}^{\dot{\alpha}}$
  and $\Pi_{\gamma}^{\alpha} \Pi_{\beta}^{\gamma} = \Pi_{\beta}^{\alpha}$.
  These vertex operators commute with the BRST charge $Q$ \eqref{q_charge} as well (still disregarding the $\vec{g}$, $\vec{h}$, $\gamma$, and $\beta$ ghosts).
  The projection operators can be defined as follows:
  Find reference spinors $s^{\alpha}$ such that $\braket{\rho_i \, s} \!\ne\! 0 \, (\forall V_i)$ and $\tilde{s}^{\dot{\alpha}}$ such that $[\tilde{\rho}_j \, \tilde{s}] \!\ne\! 0 \, (\forall \tilde{V}_j)$ 
  \footnote{If such a reference spinor $s$ or $\tilde{s}$ cannot be found, one can always deform the $\rho_i$ or $\tilde{\rho}_i$ a little bit to ensure the existence of the
  reference spinor(s) and take the limit at the end.} and define
  \begin{equation}
  \Pi_{(\rho_i) \alpha}^{\beta} = \frac{\rho_{i \, \alpha} s^{\beta}}{\braket{\rho_i \, s}}, \,\,\, 
  \tilde{\Pi}_{(\tilde{\rho}_j) \dot{\alpha}}^{\dot{\beta}} = \frac{\tilde{\rho}_{j \, \dot{\alpha}} \tilde{s}^{\dot{\beta}}}{[\tilde{\rho}_j \, \tilde{s}]}.
  \label{ProjectionOps}
  \end{equation}
  It is important to note that in scattering amplitudes one should not perform contractions of ghost fields between $V_i$ and $\tilde{V}_j$, i.e. between $E_j$ and $\tilde{F}_i$
  or between $\tilde{E}_i$ and $F_j$ fields, because such contractions would involve ghosts which were considered as decoupled in the integrated vertex 
  operators \eqref{GRvertex}. This can also be seen by using the projection operators \eqref{ProjectionOps} in cross-contractions resulting in a factor of 
  $\braket{\rho_j s}$ for some $\tilde{V}_j$ or $[\tilde{\rho}_i \tilde{s}]$ for some $V_i$ in some summand to the scattering amplitude and for such a summand a possible choice
  for the reference spinor is $s = \rho_j$ or $\tilde{s} = \tilde{\rho}_i$, making the summand vanish.\\
    
  Now a N$^k$MHV scattering amplitude
  \begin{equation*}
  \mathcal{M} = \left< c \, \tilde{\mathcal{V}}_1 \prod_{i=2}^k \! \tilde{V}_j \prod_{p=k+1}^{n-1} \!\! V_p \, \mathcal{V}_n \right>
  \end{equation*}
  looks, up to a normalization factor, very similar to one of the Einstein gravity amplitudes (15) in reference \cite{Geyer:2014}, with $E, F, \tilde{E}, \tilde{F}$ replacing the auxiliary fields
  $\rho, \tilde{\rho}$ there and without the external supermomenta. The projection operators $\Pi_{(\rho_i\!)}$ and $\tilde{\Pi}_{(\tilde{\rho}_i\!)}$ in the vertex operators have the consequence that the 
  contractions among the $E$ and $F$ fields on the one side and among the $\tilde{E}$ and the $\tilde{F}$ fields on the other side become one-dimensional
  (any cycle of contractions will end up with the trace of a product of projection operators which is always 1) and, therefore,
  lead to the same outcome as the contractions between the $\rho, \tilde{\rho}$ fields,
  such that the model produces the same tree amplitudes as Skinner's model \cite{Skinner:2013} (formula (18) in \cite{Geyer:2014}). One major difference though is that our model would produce quite different
  amplitudes when external fermions were included because they would lead to contributions from their momenta beyond mere insertions in delta functions, due to 
  contractions between bosonic ghosts that would become part of the vertex operators.\\
  
  Next we look at one-loop scattering amplitudes on the torus \cite{Lipstein:2015, Adamo:2013}. The solutions to the scattering equations and Wick contractions involve the torus Green's functions \cite{Itoyama:2015}.
  For even spin structures of type $\alpha \in \{2,3,4\} $:
  \begin{equation}
  \begin{split}
  \lambda(z) &= \sum_{i = 1}^k t_i \rho_i S_{\alpha}(z, z_i; \tau),\\
  \tilde{\lambda}(z) &= \sum_{j = k+1}^n \!\!  t_j \tilde{\rho}_j S_{\alpha}(z, z_j; \tau) \\
  S_{\alpha} (z, z_i; \tau) & = \frac{\theta_{\alpha} (z \! - \! z_i; \tau)} {\theta _{\alpha} (0; \tau)} \frac{\theta_1^{'}\!(0; \tau)}{\theta_1(z \! - \! z_i; \tau)},\\ 
  \end{split}  
  \label{even_spin}
  \end{equation}
  and for the odd spin structure
  \begin{equation}
  \begin{split}
  \lambda(z) &= \lambda_0 + \sum_{i = 1}^k t_i \rho_i S_1(z, z_i; \tau),\\
  \tilde{\lambda}(z) &= \tilde{\lambda}_0 + \sum_{j = k+1}^n \!\!  t_j \tilde{\rho}_j S_1(z ,z_j; \tau) \\
  S_1(z, z_i; \tau) & =  \frac{\theta_1^{'}\!(z - z_i; \tau)}{\theta_1(z \! - \! z_i; \tau)} - 4 \pi \frac{\mathrm{Im}(z \! - \! z_i)}{\mathrm{Im}(\tau)} \\
  \end{split} 
  \label{odd_spin}
  \end{equation}
  
  First we restrict ourselves to even spin structures only, i.e. the NS sector of the twistors, where in contrast to \cite{Lipstein:2015} there are no zero modes of $\lambda$ and 
  $\tilde{\lambda}$.
  We use the same vertex operators as before, 
  except that only integrated vertex operators appear in the amplitude at genus 1, if we take care of the one zero mode of the $c$ ghost by dividing over volGL(1). 
  The result looks similar to, but differs somewhat from \cite{Lipstein:2015}:
  
  \begin{equation}
  \begin{split}
  \mathcal{M} = & \int \!\! \frac{\mathrm{d}\tau}{\mathrm{volGL}(\!1\!) \, \mathrm{Im}\tau} \prod_{i=1}^n  \frac{\mathrm{d} z_i \mathrm{d} t_i} {t_i^3}\,\,
  \prod_{l=1}^k \!\ \delta^2(\tilde{\rho}_l \! - \! t_l \tilde{\lambda}(z_l)) \!\!\prod_{r=k+1}^n \!\!\! \delta^2(\rho_r \! - \! t_r \lambda(z_r))\\
  &\sum_{\alpha = 2,3,4} \!\! \mathrm{det} H_{\alpha} \mathrm{det} \tilde{H}_{\alpha}\, \left( \mathrm{Im}\tau (\eta(\tau))^4\right)^{-6},
  \end{split}
  \label{loop_ampl}
  \end{equation}
  where  $H_{\alpha}$ is a $k \times k$ matrix and $\tilde{H}_{\alpha}$ is an $n \!-\! k \times n \!-\! k$ matrix
  arising from the Wick contractions between the matter and ghost fields appearing in the 
  vertex operators with the following non-zero elements \footnote{Note that the (non-spinor) ghost fields are Ramond and periodic and contract to $S_1$ for any spin structure.}:
  
  \begin{equation*}
  \begin{split}
  &H_{\alpha}^{lm} = t_l t_m \braket{\rho_l \rho_m } S_1(z_l, z_m; \tau), \, l \ne m,\\
  &H_{\alpha}^{\,ll} = - \sum_{m \neq l} t_l t_m \braket{\rho_l \rho_m } S_{\alpha}(z_l, z_m; \tau)
  \end{split}
  \end{equation*}
  for $l,m \in \{1, \cdots, k\}$, and
  \begin{equation*}
  \begin{split}
  &\tilde{H}_{\alpha}^{rs} = t_r t_s \, [\tilde{\rho}_r \tilde{\rho}_s] \, S_1(z_r, z_s; \tau), \, r \ne s,\\
  &\tilde{H}_{\alpha}^{rr} = - \sum_{s \neq r} t_r t_s [\tilde{\rho}_r \tilde{\rho}_s] S_{\alpha}(z_r, z_s; \tau)
  \end{split}
  \end{equation*}
  for $r,s \in \{k+1, \cdots, n\}$, $\eta(\tau)$ is the Dedekind eta function, 
  and $\mathrm{Im}\tau^{-7}( \eta(\tau) )^{-24}$ is the contribution from the one-loop partition function, where the matter fields and some of the ghosts cancel each other and we are left with a 
  contribution from the b-c system including zero modes and 13 bosonic ghost  fields and their antighosts
  \footnote{This includes a factor of $\sqrt{\mathrm{Im} \tau}$ extracted from volGL(1) which, therefore, is to be taken as modular invariant.}.
  Under modular transformations the Green's functions $S_{\alpha}$ for $\alpha = 2,3,4$ (when multiplied with 2 $t$'s) permute among each other, whereas $S_1$ (when multiplied with 2 $t$'s) and the contribution from the one-loop partition function themselves are invariant, such that the whole scattering amplitude is modular invariant without any need for GSO-type projection. That the model has just the right number 12 of remaining bosonic ghost-antighost pairs to achieve modular invariance seems non-trivial but, of course, is connected to the zero central charge of the Virasoro algebra.\\
  
  The amplitude seems to be IR divergent in $q = e^{2i\pi\tau}$ for $q \rightarrow 0$. First it looks as if the IR divergence is as bad as for bosonic strings because of the 
  $q^{-2}\mathrm{d}q$ behavior
  of $( \eta(\tau) )^{-24}\mathrm{d}\tau$ for $q \rightarrow 0$, but actually the IR singularity gets cancelled due to the behavior of the torus Green's functions in that limit:
  
  \begin{equation}
   S_{1, 2}(z, w; \tau) \sim \frac{1}{z - w}, \,\,\, S_{3,4} (z, w; \tau) \sim \text{const},
   \label{GreenAsymptotics}
  \end{equation}
  
  where on the right, in abuse of notation, affine coordinates on the Riemann sphere have been chosen \cite{Adamo:2013}.
  Namely, from this it follows that for $q \rightarrow 0$ not only the matrices $H_2$ and $\tilde{H}_2$ have the same structure as in the tree scattering case on the Riemann sphere 
  and, therefore, have co-rank 1 with determinant vanishing each as $q$ for $q \rightarrow 0$, but also that the
  matrices $H_3$ and $H_4$ are the same in this limit on the support of scattering equations \eqref{even_spin}
  that develop zero modes, which is not possible for even spin structures such that
  the determinants of $H_3$ and $H_4$ also must vanish as $q$ for $q \rightarrow 0$, and the same for $\tilde{H}_3$ and $\tilde{H}_4$.
  Whether the full or at least the leading IR singularity cancels for other Riemann surfaces and multiple loops as well, remains to be seen.
  In any case one can argue that IR-divergences can be dealt with in a systematic fashion once one has a second quantized version
  of the model, i.e. a string field theory \cite{Sen:2015}.\\
  
  Let's look now at odd spin structures.
  In this case amplitudes are in the R sector and the zero modes of the matter fields need to be added to the solutions of the scattering equations 
  and integrated over \cite{Lipstein:2015, Adamo:2013}. The integration over the zero modes $\mu_0$ and $\tilde{\mu}_0$ adds a momentum conserving delta function.
  Otherwise the amplitude looks similar to \eqref{loop_ampl} with additional integration over the zero modes $\lambda_0$ and $\tilde{\lambda}_0$,
  and the sum over $\mathrm{det}H_{\alpha} \mathrm{det}\tilde{H}_{\alpha}$ replaced by a single $\mathrm{det}H_1 \mathrm{det}\tilde{H}_1$
  which has the same off-diagonal elements but different diagonal ones:
  
  \begin{equation*}
  \begin{split}
  \mathcal{M} = & \delta^4(\sum_{i=1}^n \rho_i \tilde{\rho}_i) \int \! \frac{\mathrm{d}\tau \, \mathrm{d}^2\lambda_0 \mathrm{d}^2\tilde{\lambda}_0}{\text{volGL}(\!1\!)} \prod_{i=1}^n  
  \frac{\mathrm{d} z_i \mathrm{d} t_i} {t_i^3}\,\, \prod_{l=1}^k \!\ \delta^2(\tilde{\rho}_l \! - \! t_l \tilde{\lambda}(z_l)) \!\!\prod_{r=k+1}^n \!\!\! \delta^2(\rho_r \! - \! t_r \lambda(z_r)) \\
  &\mathrm{det} H_1\mathrm{det} \tilde{H}_1 \left( \mathrm{Im}\tau (\eta(\tau))^4\right)^{-6},
  \end{split}
  \end{equation*}
  where for $l,m \in \{1, \cdots, k\}$, and $r,s \in \{k+1, \cdots, n\}$:
  \begin{equation*}
  \begin{split}
  &H_1^{lm} = H_{\alpha}^{lm} = t_l t_m \braket{\rho_l \rho_m } S_1(z_l, z_m; \tau), \, l \ne m\\
  &H_1^{\,ll} = - t_l \braket{\rho_l \lambda_0} - \sum_{m \neq l} H_1^{lm} = -t_l \, \mathrm{lim}_{z \rightarrow z_l} \!\! \braket{\rho_l \lambda(z)},\\
  &\tilde{H}_1^{rs} = \tilde{H}_{\alpha}^{rs} = t_r t_s [\tilde{\rho}_r \tilde{\rho}_s] S_1(z_r, z_s; \tau), \, r \ne s\\
  &\tilde{H}_1^{rr} = - t_r [ \tilde{\rho}_r \tilde{\lambda}_0] - \sum_{s \neq r} \tilde{H}_1^{rs} = -t_r \, \mathrm{lim}_{z \rightarrow z_r} [ \tilde{\rho}_r \tilde{\lambda}(z)]
  \end{split}
  \end{equation*}
  
  Because $S_1$ is modular invariant, so is the new amplitude.
  The fact that the partition functions contribute substantially to the 
  amplitude differs from the corresponding amplitude for higher-dimensional ambitwistors in \cite{Adamo:2013}. 
  Concerning the IR-divergence, we observe that for $q \rightarrow 0$ the Green's function $S_1$ behaves as in 
  \eqref{GreenAsymptotics} leading to the same scattering equations as for genus 0 
  but for the odd spin structure there are no solutions of such scattering equations, i.e. the IR singularity must vanish like for even spin structures. \\
  
  UV-finiteness of the amplitude is not obvious because of the integration over zero modes. However one can argue that from \eqref{odd_spin} and momentum
  conservation it follows \cite{Lipstein:2015}
  \footnote{The conditions $\sum_{l=1}^k \! t_l \rho_l = 0 = \sum_{r=k+1}^n \!t_r \tilde{\rho}_r$ of \cite{Lipstein:2015} seem too strong, e.g. for $k=2$ it would follow 
  $t_1 = t_2 = 0$ in general.}
  \begin{equation*}
  \sum_{i=1}^n \rho_i \tilde{\rho}_i = \sum_{l=1}^k \! t_l \rho_l \, \tilde{\lambda}_0 + \lambda_0 \! \sum_{r=k+1}^n \!\!t_r \tilde{\rho}_r = 0
  \end{equation*}
  such that zero modes $\lambda_0$ and $\tilde{\lambda}_0$ need to become large simultaneously and then on the support of the delta functions in the integral for the 
  amplitude it follows that the $t_i$ need to stay finite and non-zero, and some of the $S_i(z_j, z_i; \tau)$ need to become large requiring that every puncture of positive 
  helicity becomes very close to at least one
  of the punctures of negative helicity and vice versa. The clusters of points close to each other can be viewed as sitting on Riemann spheres connected to the torus through very 
  large tubes and would separate in the limit which, again, is impossible for odd spin structure, i.e. the integrand must vanish at least as 
  $\prod_{\alpha} \lambda_{0 \, \alpha}^{-1} \prod_{\dot{\alpha}} \tilde{\lambda}_{0 \, \dot{\alpha}}^{-1}$ and the amplitude has at most a logarithmic UV divergence.\\
  
  In summary, the foregoing should have made it clear, that any one-loop open or closed string 
  amplitude will have a contribution from the one-loop partition function and will, therefore, be likely modular invariant.

\section{Summary and Outlook}
The model presented here has been shown to exhibit some interesting properties. In section \ref{Model} the model was defined and it was proven (with details in the appendix \ref{BRST}) that it is anomaly-free with a nilpotent BRST charge and low lying ground states in both the NS and R sector. In section \ref{VertexOps} the result was obtained that the model leads to expected gravitational N$^k$MHV tree amplitudes and to one-loop level amplitudes exhibiting modular invariance.\\

There are many open issues left:
\begin{itemize}
\item The model can include various types of strings, oriented open and closed chiral strings of either chirality and also unoriented strings and oriented closed strings.
Strings with both left and right moving modes have a large number of ground states on the same lowest level, so might lead to a too large Hilbert space.
It would be useful to investigate this more.
\item A second quantized version, e.g. along the lines of \cite {Moosavian:2019}
\footnote{In \cite{Reid:2017} arguments were given to use a similar construction for the ambitwistor string in 10 dimensions.},
is needed in order to check whether the model is consistent and unitary, and whether divergences can be removed like in superstring field theory \cite{Sen:2015}. It would also
provide a better understanding of the classical limit if it exists, via the 1PI version of the string field theory at genus 0 \cite {Moosavian:2019}.
\item Also, does the model have a low energy limit and would it link to a QFT? What would be the spectrum of particles and what would be the physical interpretation?
\end{itemize}
    
\section*{Acknowledgements}
  This paper is mostly based on other work done in the scientific community in the area of string theory and twistor theory. In particular, it benefited greatly from the contributions in the references. The authors all deserve my gratitude.\\

\appendix
\section{BRST Quantization}
  \label{BRST}
  BV quantization of the model \eqref{action} is straight forward, and in the conformal gauge that sets all Lagrange multipliers to zero, $a_{k i j} = b_{i j} = \vec{c} = 0$, the action becomes:
  
  \begin{multline}
  S_1 = \frac{1}{2 \pi} \int\mathop{}\negthickspace{\ud^2\mathop{}\negthickspace z} \left\{
  \frac{1}{2} \sum_{i=1}^2 \left( Y_i\overline\partial{Z_i} - Z_i \overline\partial{Y_i} + \Theta_i\overline\partial{\Psi_i}  + \Psi_i\overline\partial{\Theta_i} \right)
  \right.\\
  + b \overline\partial{c}
  + \sum_{i,j=1}^2 \left( \beta_{1 i j} \overline\partial{\gamma_{1 i j}} + \beta_{2 i j} \overline\partial{\gamma_{2 i j}} + f_{i j} \overline\partial{e_{i j}} \right)
  + \vec{h} \cdotp \overline\partial{\vec{g}} \left. \vphantom{\sum_i} \right\}
  \label{gf_action}
  \end{multline} 
  
  with fermionic ghost-antighost pairs $(c, b)$, $(\vec{g}, \vec{h})$, $(e_{i j}, f_{i j})\!  =\! (e_{i j}^{\dot{\alpha}_i \alpha_j}, f_{i j\, \dot{\alpha}_i \alpha_j})$  
  and bosonic ghost-antighost pairs $((\gamma_{1ij}, \beta_{1ij})\! =\! (\gamma_{1ij}^{\alpha_i \dot{\beta}_j}, \beta_{1ij\, \alpha_i \dot{\beta}_j})$, $(\gamma_{2ij}, \beta_{2ij})\! =\! (\gamma_{2ij}^{\dot{\alpha}_i \beta_j}, \beta_{2ij\, \dot{\alpha}_i \beta_j} )$ belonging to the currents (without ghosts):

  \begin{equation*}
  \begin{split}
  &T = \frac{1}{2} \sum_{i=1}^2 :\! \left( Y_i \partial{Z_i} - (\partial{Y_i}) Z_i + \Theta_i \partial{\Psi_i} - (\partial{\Theta_i}) \Psi_i \right) \!:,\\
  &\vec{H} = - :\!(Y_1 Y_2) \vec{\tau} \binom{Z_1}{Z_2}\!:,\\
  &F_{i j\,\dot{\alpha}_i \alpha_j} = - \tilde\lambda_{i \dot{\alpha}_i} \lambda_{j \alpha_j},\\  
  &G_{1i j\,\alpha_i \dot{\beta}_j} = \lambda_{i \alpha_i} \tilde\phi_{j \dot{\beta}_j},\\ 
  &G_{2i j\,\dot{\alpha}_i \beta_j} = - \tilde\lambda_{i \dot{\alpha}_i} \phi_{j \beta_j}.
  \end{split}
  \end{equation*}
  \\
  Using the mode expansions
  
  \begin{align*}
  Z_i & = \sum_n \! Z_{i\,n} z^{-n-\frac{1}{2}}\,,\, & Y_i & = \sum_n \! Y_{i\,n} z^{-n-\frac{1}{2}}\,, &\\
  \Psi_i & = \sum_n \! \Psi_{i\,n} z^{-n-\frac{1}{2}}\,, & \Theta_i & = \sum_n \! \Theta_{i\,n} z^{-n-\frac{1}{2}}, &\\
  c & = \sum_n \! c_n z^{-n+1}\,,\, & b & = \sum_n b_n z^{-n-2}\,, &\\
  \vec{g} & = \sum_n \! \vec{g}_n z^{-n}\,, & \vec{h} & = \sum_n \vec{h}_n z^{-n-1}\,, &\\
  e_{i j} & = \sum_n \! e_{i j n} z^{-n}\,,\, & f_{i j} & = \sum_n f_{i j n} z^{-n-1}\,, &\\
  \gamma_{k i j} & = \sum_n \! \gamma_{k i j n} z^{-n}\,,\, & \beta_{k i j} & = \sum_n \beta_{k i j n} z^{-n-1}\,, &
  \end{align*}
  
  with $n \, \epsilon \, \mathbb{Z}$ for Ramond (R) and $n \, \epsilon \, \mathbb{Z}\!+\!\frac{1}{2}$ for Neveu-Schwarz (NS) boundary conditions, we get for the left-moving currents (without ghosts)
  \begin{equation*}
  T = \sum L_n z^{-n-2}, \vec{H} = \sum \vec{H}_n z^{-n-1}, F_{i j} = \sum F_{i j\, n} z^{-n-1}, G_{k i j} = \sum G_{k i j\, n} z^{-n-1}
  \end{equation*}
  with
  \begin{equation}
  \begin{split}
  & L_n = \sum_{i \, m} \left(- \frac{n}{2} + m \right) :\! \left( Y_{i\,n-m} Z_{i\,m} + \Theta_{i\,n-m} \Psi_{i\,m} \right) \!:,\\
  & \vec{H}_n = -\sum_m :\!(Y_1 Y_2)_{n-m} \, \vec{\tau}\, {\binom{Z_1}{Z_2}}_m \!\!\! :,\\
  & F_{i j \, n} = - \sum_m \tilde\lambda_{i\, n-m} \lambda_{j \, m},\\
  & G_{1 i j \, n} = \sum_m \lambda_{i \, n-m} \tilde\phi_{j \, m},\\
  & G_{2 i j \, n} = - \sum_m \tilde\lambda_{i \, n-m} \phi_{j \, m}.
  \end{split}
  \label{currents}
  \end{equation}
  
  Here
  \begin{equation}
  \begin{aligned}
  &(\tilde{\mu}_{i\,-n}, \! \lambda_{i\,n}),  (\lambda_{i\,-n}, \! \tilde{\mu}_{i\,n}),  (\tilde{\lambda}_{i\,-n}, \! \mu_{i\,n}),  (\mu_{i\,-n}, \! \tilde{\lambda}_{i\,n}), 
  (\tilde{\psi}_{i\,-n}, \! \phi_{i\,n}),  (\phi_{i\,-n}, \! \tilde{\psi}_{i\,n}), (\tilde{\phi}_{i\,-n}, \! \psi_{i\,n}), (\psi_{i\,-n}, \! \tilde{\phi}_{i\,n}), \\
  &(c_{-n+2}, \! b_{n-2}), (b_{-n-1}, \! c_{n+1}), (\vec{g}_{-n}, \! \vec{h}_n), (\vec{h}_{-n}, \! \vec{g}_n),
  (e_{ij\,-n}, \! f_{ij\,n}), (f_{ij\,-n}, \! e_{ij\,n}), \\
  &(\gamma_{kij\,-n}, \! \beta_{kij\,n}), (\beta_{kij\,-n}, \! \gamma_{kij\,n})
  \end{aligned}
  \label{modes}
  \end{equation}
  are considered creation-annihilation operator pairs for $n > 0$ and
  \begin{equation*}
  \begin{aligned}
  &(\tilde{\mu}_{i\,0}, \lambda_{i\,0}), (\mu_{i\,0}, \tilde{\lambda}_{i\,0}), (\tilde{\psi}_{i\,0}, \phi_{i\,0}), (\psi_{i\,0}, \tilde{\phi}_{i\,0}),
  (\vec{g}_{0}, \vec{h}_0), (e_{ij\,0}, f_{ij\,0}), (\gamma_{kij\,0}, \beta_{kij\,0})
  \end{aligned}
  \end{equation*}  
  creation-annihilation operator pairs for $n = 0$ (in the R sector with picture number $\mp1/2$ for bosonic/fermionic modes and in the NS sector with picture number $\mp1$). \\
  
  The currents form an algebra with commutators fulfilling the Jacobi identity: \footnote{The central charge for Virasoro algebra is zero because of same number of bosons and fermions.}
  \begin{equation}
  \begin{split}
  & [ L_n, L_m ] = (n - m)L_{n+m} \\
  & [ G_{^1_2 i j \, n} , L_m ] = n \, G_{^1_2 i j \, n+m} \\
  & [ F_{i j \, n} , L_m ] = n \, F_{i j \, n+m} \\ 
  & [ \vec{H}_n , L_m ] = n \, \vec{H}_{n+m} \\ 
  & \{ G_{^1_2 i j \, n} , G_{^1_2 k l \, m} \} = 0 \\ 
  & \{ G_{1 i j \, n} , G_{2 k l \, m} \} = 0 \\
  & [ G_{^1_2 i j \, n} , F_{k l \, m} ] = 0 \\ 
  & [ H_n^3,  G_{^1_2 i j \, m} ] = \, \mp \frac{1}{2} (-1)^i G_{^1_2 i j \, n+m} \\  
  & [ H_n^+,  G_{^1_2 i j \, m} ] = \, \pm \delta_{i^1_2} G_{^1_2 3-i \, j \, n+m} \\  
  & [ H_n^-,  G_{^1_2 i j \, m} ] = \, \pm \delta_{i^2_1} G_{^1_2 3-i \, j \, n+m} \\  
  & [ F_{i j \, n} , F_{k l \, m} ] = 0 \\     
  & [ H_n^3,  F_{i j \, m} ] = (-1)^i \delta_{3\,i+j} F_{i j \, n+m} \\ 
  & [ H_n^+,  F_{i j \, m} ] = \delta_{j1} F_{i 2 \, n+m} - \delta_{i2} F_{1 j \, n+m} \\   & [ H_n^-,  F_{i j \, m} ] = \delta_{j2} F_{i 1 \, n+m} - \delta_{i1} F_{2 j \, n+m} \\ 
  & [ H_n^i,  H_m^i ] = \, n \delta_{3i} \delta_{0\,n+m} \\ 
  & [ H_n^+,  H_m^- ] = - 2 H^3_{n+m} + 2n\delta_{0\,n+m} \\ 
  & [ H_n^\pm,  H_m^3 ] = \pm H^\pm_{n+m} 
  \end{split}
  \label{commut}
  \end{equation}
  
  The only anomalies arise from the commutators of the rotational generators $\vec{H}$ with themselves.
  We will show that after inclusion of ghosts new anomalies arise, but eventually all anomalies will cancel each other.\\[1pt]
  
  The BRST charge operator $Q$ is obtained from the commutators $[K_i, K_j] = f_{ij}^k K_k$ using the formula $Q = c^i K_i - \frac{1}{2} f^{ij}_k c^i c^j b_k$, where $c^i$ denote the ghosts and $b_k$ the antighosts
  (see e.g. 3.2.1 in \negthickspace ~\cite {GSW:1987}).
  In our case $Q$ becomes:  
  \begin{multline}  
  Q = \sum c_{-n} L_n + \sum \gamma^{\alpha_i \dot{\beta}_j}_{1 i j \, -n} G_{1i j\,\alpha_i \dot{\beta}_j \, n} + \sum \gamma^{\dot{\alpha}_i \beta_j}_{2 i j \, -n} G_{2i j\,\dot{\alpha}_i \beta_j \, n} + \sum e_{i j \, -n}^{\dot{\alpha}_i \alpha_j} F_{i j\,\dot{\alpha}_i \alpha_j \, n} \\
  + \sum \vec{g}_{-n} \vec{H}_n - a \, c_0 - \frac{1}{2} \sum  (n-m) \, :c_{-n} c_{-m} b_{m+n} :\\
  - \sum m c_{-n} : \gamma^{\alpha_i \dot{\beta}_j}_{1 i j \, -m} \beta_{1ij\, \alpha_i \dot{\beta}_j  \, m+n} :
  - \sum m c_{-n} : \gamma^{\dot{\alpha}_i \beta_j}_{2 i j \, -m} \beta_{2ij\, \dot{\alpha}_i \beta_j  \, m+n} : \\
  + \sum m c_{-n} : e_{i j \, -m}^{\dot{\alpha}_i \alpha_j} f_{i j\, \dot{\alpha}_i \alpha_j \, m+n} : + \sum m c_{-n} : \vec{g}_{-m} \vec{h}_{m+n} :
  + \frac{i}{2} \sum : \vec{g}_{-n} \left( \vec{g}_{-m} \times \vec{h}_{m+n} \right) : \\
  + \sum \vec{g}_{-n} : \left[ 
  \left(\gamma_{11j} \gamma_{12j} \right)_{-m} \mathop{}\negthickspace \vec{\tau}  \mathop{}\negthickspace \begin{pmatrix} \beta_{11j} \\ \beta_{12j} \end{pmatrix}_{\! m+n} 
   - \left(\beta_{21j} \beta_{22j} \right)_{-m} \mathop{}\negthickspace \vec{\tau}  \mathop{}\negthickspace \begin{pmatrix} \gamma_{21j} \\ \gamma_{22j} \end{pmatrix}_{\! m+n} \right. \\ 
  - \left(e_{11} e_{12} \right)_{-m} \mathop{}\negthickspace \vec{\tau}  \mathop{}\negthickspace \begin{pmatrix} f_{11} \\ f_{12} \end{pmatrix}_{\! m+n} 
  - \left(e_{21} e_{22} \right)_{-m} \mathop{}\negthickspace \vec{\tau}  \mathop{}\negthickspace \begin{pmatrix} f_{21} \\ f_{22} \end{pmatrix}_{\! m+n}\\
  - \left(f_{11} f_{21} \right)_{-m} \mathop{}\negthickspace \vec{\tau}  \mathop{}\negthickspace \begin{pmatrix} e_{11} \\ e_{21} \end{pmatrix}_{\! m+n}
  - \left(f_{12} f_{22} \right)_{-m} \mathop{}\negthickspace \vec{\tau}  \mathop{}\negthickspace \begin{pmatrix} e_{12} \\ e_{22} \end{pmatrix}_{\! m+n}
  \left. \vphantom{\begin{pmatrix} \beta_{21j} \\ \beta_{22j} \end{pmatrix}_{\! m+n}}\right] :
  \label{q_charge}
   \end{multline} 
   
   where summation is assumed over any repeated indices.\\[1pt]
   
   The full current algebra $L_n^{tot} \! = \! \{Q,b_n\}$, $F_{ij}^{tot} \! = \! \{Q, f_{ij\,n}\}$, $G_{kij}^{tot} \! = \! [Q, \beta_{kij\,n}]$, and $\vec{H}^{tot} \! = \! \{Q, \vec{h}_n\}$ including the ghosts has the same commutator relations as in \eqref{commut}, but without any anomalies, as can be seen from follows:
   
   Concerning the Virasoro algebra, besides the $(b,c)$ system with conformal weight $(2,0)$ and $(-1,0)$ contributing $\frac{n}{6}(1 - 13n^2)$ to the central charge, we have 32 
   bosonic and 19 fermionic
   antighost-ghost pairs of conformal weight $(1, 0)$ and $(0, 0)$ contributing to the central charge $\frac{n}{6}(13n^2 - 1 + \alpha)$, where $\alpha$ depends on which sectors,
   NS or R, are considered.
   When all fields are in the R sector, then $\alpha = -12$ and setting $a=1$ in the equation
   \eqref{q_charge}
   for $Q$ ensures vanishing central charge of the Virasoro algebra. When matter fields are changed to the NS sector, \eqref{q_charge} requires that the ghosts and antighosts are kept R, and $\alpha$ and $a$ obtain the same value. When fermionic and bosonic matter are in different sectors, then the matter fields have a non-zero contribution to $\alpha$, the 32 fields $(\beta,\gamma)$ must change to NS, and $a$ takes on a negative value $-2$ (R twistors and NS fermionic spinors) or $-4$ (NS twistors and R fermionic spinors), which does not seem to have any solutions and will be discarded as uninteresting (it would also break the fermionic symmetries). Similarly (negative $a$ and fermionic symmetry breaking) for the case, when half of bosonic matter is in different sectors (this also breaks $SU(2)$) and/or half of fermionic matter is in different sectors.
   In any case, the full Virasoro algebra can be made anomaly-free with an appropriate value of $a$, taken to be equal to 1 from now on and, therefore, with all matter fields either 
   in the NS or in the R sector.
   
   Of the remaining currents only the total SU(2) currents $\vec{H}_n^{tot}$ could develop anomalies. Actually it is straight forward to see that in the commutators between
   components of $\vec{H}_n^{tot}$ the anomalies arising from $(\vec{h},\vec{g})$ cancel against the anomalies from the ghost-free currents $\vec{H}$ and the 
   anomalies from the bosonic $(\beta, \gamma)$ cancel against the anomalies from $(e,f)$, i.e. the $\vec{H}_n^{tot}$ generate an anomaly-free algebra. This can also
   easily be seen by computing the anomaly coefficient in terms of the Casimir $C_2(F)=3/4$ of the (anti-)fundamental representation $F$ and $C_2(A)=2$ of the 
   adjoint representation $A$ of SU(2):\\\
   $4 \times 2 C_2(F) - 3 C_2(A) + 16 \times 2 C_2(F) - 16 \times 2 C_2(F) = 0$ \footnote{The four $e\!-\!f$ terms in the $H_3^{tot}$ component are unique in the sense that some parts 
   of the terms cancel against each other and some add up, such that the final contribution to the anomaly is the same as for the other components.}. \\[1pt]
   
   Therefore, for $a = 1$ $Q$ is nilpotent, $Q^2 = 0$,  and anomaly-free as well.\\[1pt]
   
   Physical states are states in the cohomology of $Q$ and are annihilated by all currents in
   \eqref{currents}
   for $n >= 0$ except for $L_0$, which has eigenvalue $a = 1$ for physical states.\\
   
   Such (single string) states include, when applied to the vacuum state, in the
   NS sector any product of two (different) creation operators out of
   $\{\!\lambda_{i\,-\frac{1}{2}}, \!\tilde{\lambda}_{j\,-\frac{1}{2}}, \!\phi_{k\,-\frac{1}{2}}, \!\tilde{\phi}_{l\,-\frac{1}{2}}\!\}$, and in the
   R sector any product of a creation operator out of  
   $\{\lambda_{i\,-1}, \tilde{\lambda}_{j\,-1}, \phi_{k\,-1}, \tilde{\phi}_{l\,-1}\}$
   with any (single) or none operator out of $\{\mu_{m\,0}, \tilde{\mu}_{n\,0}, \psi_{r\,0}, \tilde{\psi}_{s\,0}\}$, except for any combination out of
   $\{\lambda_{i\,-1} \tilde{\mu}_{n\,0}, \tilde{\lambda}_{j\,-1} \mu_{m\,0}, \phi_{k\,-1} \tilde{\psi}_{s\,0} , \tilde{\phi}_{l\,-1} \psi_{r\,0}\}$ (not annihilated by $L_1$), 
   whereby resulting states containing bosonic spinors need to be superposed or projected such that the resulting state is $SU(2)$-neutral (annihilated by $\vec{H}_n, n \geq 0$).\\
   
   All this is complemented by the corresponding right-moving entities.  
  
\appendix

\bibliography{TwistorString}

\begin{thebibliography}{10}

\bibitem{Penrose:1986}
R.~Penrose and W.~Rindler, {\em Spinors and space-time, Volume 2}.
\newblock Cambridge University Press, 1986.

\bibitem{LeBrun:1991}
C.~LeBrun, ``Thickenings and conformal gravity,'' {\em Commun. Math. Phys.},
  vol.~139, pp.~1--43, 1991.

\bibitem{Witten:2004}
E.~Witten, ``Perturbative gauge theory as a string theory in twistor space,''
  {\em arXiv:0312171v2 [hep-th]}, 2004.

\bibitem{Skinner:2013}
D.~Skinner, ``Twistor strings for n = 8 supergravity,'' {\em
  arXiv:1301.0868[hep-th]}, 2013.

\bibitem{Geyer:2014}
Y.~Geyer, A.~E. Lipstein, and L.~Mason, ``Ambitwistor strings in
  4-dimensions,'' {\em arXiv:1404.6219v2[hep-th]}, 2014.

\bibitem{Polyakov:2005}
D.~Polyakov, ``Conformal moduli and b - c pictures for nsr strings,'' {\em
  arXiv:0401009v3[hep-th]}, 2004.

\bibitem{Lipstein:2015}
A.~Lipstein and V.~Schomerus, ``Towards a worldsheet description of $n = 8$
  supergravity,'' {\em arXiv:1507.02936[hep-th]}, 2015.

\bibitem{Adamo:2013}
T.~Adamo, E.~Casali, and D.~Skinner, ``Ambitwistor strings and the scattering
  equations at one loop,'' {\em arXiv:1312.3828v2[hep-th]}, 2013.

\bibitem{Itoyama:2015}
H.~Itoyama and K.~Yano, ``Genus one super-green function revisited and
  superstring amplitudes with non-maximal supersymmetry,'' {\em
  arXiv:1512.07705[hep-th]}, 2015.

\bibitem{Sen:2015}
A.~Sen, ``Ultraviolet and infrared divergences in superstring theory,'' {\em
  arXiv:1512.00026[hep-th]}, 2015.

\bibitem{Moosavian:2019}
S.~F. Moosavian, A.Sen, and M.~Verma, ``Superstring field theory with open and
  closed strings,'' {\em arXiv:1907.10632v1[hep-th]}, 2019.

\bibitem{Reid:2017}
R.~A. Reid-Edwards and D.~A. Riccombeni, ``A superstring field theory for
  supergravity,'' {\em arXiv:1701.05495v3 [hep-th]}, 2017.

\bibitem{GSW:1987}
M.~Green, J.~Schwarz, and E.~Witten, {\em Superstring theory, Volume 1}.
\newblock Cambridge University Press, 1987.

\end{thebibliography}
\end{document}